\documentclass[prd,onecolumn]{revtex4}
\usepackage{dcolumn}
\usepackage{multirow}
\usepackage{graphicx}
\usepackage{amssymb}
\usepackage{bm}
\usepackage{hyperref}
\usepackage{epstopdf}
\usepackage{color}
\usepackage{mathrsfs}
\usepackage{amsmath,amssymb,amsthm}
\usepackage{rotating}
\usepackage{sverb, longtable}
\usepackage{subfigure}

\usepackage[graphicx]{realboxes}
\usepackage{adjustbox}
\begin{document}

\title{Can $f(R)$ gravity relieve $H_0$ and $\sigma_8$ tensions? }
\author{Deng Wang}
\email{cstar@sjtu.edu.cn}
\affiliation{National Astronomical Observatories, Chinese Academy of Sciences, Beijing, 100012, China}
\begin{abstract}
To investigate whether $f(R)$ gravity can relieve current $H_0$ and $\sigma_8$ tensions, we constrain the Hu-Sawicki $f(R)$ gravity with Planck-2018 cosmic microwave background and redshift space distortions observations. We find that this model fails to relieve both $H_0$ and $\sigma_8$ tensions, and that its two typical parameters  
$\log_{10}f_{R0}$ and $n$ are insensitive to other cosmological parameters. Combining the cosmic microwave background, baryon acoustic oscillations, Type Ia supernovae, cosmic chronometers with redshift space distortions observations, we give our best constraint $\log_{10}f_{R0}<-6.75$ at the $2\sigma$ confidence level.

\end{abstract}
\maketitle

\section{Introduction}
Since the accelerated expansion of the universe is discovered by Type Ia supernovae (SNe Ia) \cite{1,2} and ensured by two independent probes baryon acoustic oscillations (BAO) \cite{3,4} and cosmic microwave background (CMB) radiation \cite{5,6,7}, the elegant standard six-parameter cosmological model, $\Lambda$-cold dark matter ($\Lambda$CDM) has achieved great success in explaining the physical phenomena at both small and large scales. Up to now, the nature of both dark energy (DE) and dark matter (DM) is still mysterious and unclear, and we just know phenomenologically the following several basic properties of them: (i) DE is a cosmic fluid with an effective equation of state (EoS) $\omega\approx-1$, which violates the strong energy condition; (ii) DE obeys a too much smaller clustering property than DM and is homogeneously permeated in the universe at cosmological scale; (iii) effects of DM clustering have been measured to 2$\sim$3$\%$ precision by several large weak lensing experiments including the Kilo-Degree Survey (KiDS) \cite{8}, the Dark Energy Survey (DES) \cite{9} and the Subaru Hyper-Suprime Camera (HSC) \cite{10}. Recently, the Planck-2018 CMB final release \cite{5} with improved measurement of the reionized optical depth has confirmed, once again, the validity of the simple $\Lambda$CDM cosmology in describing the evolution of the universe. However, this model is not as perfect as we imagine and faces at least two intractable problems, namely the cosmological constant and coincidence problems. The former indicates that the observed value for vacuum energy density is far smaller than its theoretical estimation, i.e., the so-called 120-orders-of-magnitude inconsistence that makes the physical explanation of vacuum very confusing, while the latter is why energy densities of DE and DM are of the same order of magnitude today, since their energy densities are so different from each other during the evolution of the universe. Meanwhile, the $\Lambda$CDM model also faces at least two important tensions emerged from recent cosmological observations, namely the Hubble constant ($H_0$) and matter fluctuation amplitude ($\sigma_8$) tensions, where the former is more severe than the latter. The $H_0$ tension is that the direct measurement of today's cosmic expansion rate from the Hubble Space Telescope (HST) is over 4$\sigma$ level higher than the indirectly derived value from the Planck-2018 CMB measurement, while the $\sigma_8$ one indicates that today's matter fluctuation amplitude in linear regime measured by several low redshift probes including weak gravitational lensing \cite{11}, cluster counts \cite{12} and redshift space distortions \cite{13} is still lower than that indirectly measured by the Planck-2018 CMB data \cite{5}. It is nature that one may query the correctness of $\Lambda$CDM in characterizing the background evolution and structure formation of the universe. As a consequence, a wide variety of cosmological models based on some physical mechanism have been proposed to explain the late-time cosmic acceleration. Most recently, due to severer $H_0$ tension and richer data from large scale galaxy survey than before \cite{14}, cosmologists have a stronger motivation and more interests to resolve or even solve these tensions by confronting existing cosmological models or constructing new ones with current observations. It is worth noting that possible systematic errors or independent determinations on $H_0$ and $\sigma_8$ from new probes can also alleviate these tensions. To resolve $H_0$ and $\sigma_8$ tensions, in previous works, many authors always combine CMB data with BAO, SNe Ia, local $H_0$ observation to give tight constraints on a specific model. We argue that, more or less, this kind of constraint can only give an indirect answer for cosmological tensions, and that the most direct method is to check the model dependence of Planck-2018 CMB data.              

In this study, our motivation is to explore whether one of the simplest extensions of general relativity (GR), $f(R)$ gravity \cite{15,16}, can relieve current $H_0$ and $\sigma_8$ tensions. In $f(R)$ gravity, the modified Friedmann equations can be obtained by varying a generalized Lagrangian which is a function of the Ricci scalar $R$. Although many authors have constrained specific $f(R)$ models with joint cosmological observations in recent years, there is still a lack of a direct test of the ability to alleviate $H_0$ and $\sigma_8$ tensions for $f(R)$ gravity in light of Planck CMB data. Especially, due to three reasons: (i) the data of Planck-2018 full mission is released; (ii) $H_0$ tension becomes more serious than before; (iii) richer data from large scale galaxy survey to study DM clustering is gradually obtained, this is an urgent issue needed to be addressed. By implementing numerical analysis, we find that the Hu-Sawicki $f(R)$ gravity cannot reduce $H_0$ and $\sigma_8$ tensions.

This work is organized as follows. In the next section, we introduce the basic equations of $f(R)$ gravity and a specific $f(R)$ model to be investigated in this analysis. In Section III, we display the data and analysis method. In Section IV, the numerical results are presented. The  discussions and conclusions are exhibited in the final section.    

\section{$f(R)$ gravity}
To construct a modified theory of gravity, one can introduce some terms such as $R^2$, $R^{\mu\nu}R_{\mu\nu}$, $R^{\mu\nu\alpha\beta}R_{\mu\nu\alpha\beta}$, or $R\square^nR$, when quantum corrections are taken into account. In $f(R)$ gravity, different from the above high-order derivative gravity, the modification is just a function of Ricci scalar $R$. $f(R)$ gravity was firstly introduced by Buchdahl \cite{15} in 1970 and the readers can find more details in recent reviews \cite{16,17}. The action is written as 
\begin{equation}
S=\int d^4x\sqrt{-g}\left[R+f(R)+\mathcal{L}_m\right], \label{1}
\end{equation}
where $f(R)$, $\mathcal{L}_m$ and $g$ denote a function of $R$, the standard matter Lagrangian and the trace of the metric, respectively. By varying Eq.(\ref{1}), one can obtain the modified Einstein field equation
\begin{equation}
G_{\mu\nu}+f_RR_{\mu\nu}+(\Box f_R-\frac{f}{2})g_{\mu\nu}-\nabla_\mu\nabla_\nu f_R=8\pi GT_{\mu\nu}, \label{2}
\end{equation} 
where $f_R\equiv df/dR$ denotes an extra scalar degree of freedom, i.e., the so-called scalaron and $T_{\mu\nu}$ is energy-momentum tensor. In a spatially flat Friedmann-Robertson-Walker (FRW) universe, the equation of background evolution in $f(R)$ gravity is expressed as 
\begin{equation}
H^2+\frac{f}{6}-(H^2+H\frac{dH}{dN})f_R+H^2\frac{dR}{dN}f_{RR}=\frac{8\pi G}{3} \rho_m,              \label{3}
\end{equation} 
where $f_{RR}\equiv df_R/dR$, $N\equiv \mathrm{ln}\,a$, $H$ is Hubble parameter, $a$ is scale factor and $\rho_m$ is matter energy density. 

We are also of interests to study the perturbations in $f(R)$ gravity and just consider the linear part here. For sub-horizon modes ($k\gtrsim aH$) in the quasi-static approximation, the linear growth of matter density perturbations is shown as \cite{18}
\begin{equation}
\frac{\mathrm{d}^2\delta}{\mathrm{d}a^2}+\left(\frac{1}{H}\frac{\mathrm{d}H}{\mathrm{d}a}+\frac{3}{a}\right)\frac{\mathrm{d}\delta}{\mathrm{d}a}-\frac{3\Omega_{m}H_0^2a^{-3}}{(1+f_R)H^2}\left(\frac{1-2X}{2-3X}\right)\frac{\delta}{a^2}=0, \label{4}
\end{equation}
where $\Omega_{m}$ denotes the effective matter density ratio at present. The function $X$ has the following form
\begin{equation}
X(k,a) = -\frac{2f_{RR}}{1+f_R}\left(\frac{k}{a}\right)^2. \label{5}
\end{equation}
It is noteworthy that the function $X$ in Eq.(\ref{4}) induces a scale dependence of linear growth factor $\delta(k,a)$ in $f(R)$ gravity, when the growth factor is just a function of scale factor $a$ in GR. 

In general, a viable $f(R)$ model should be responsible for the inflationary behavior in the very early universe, reproduce the late-time cosmic acceleration, pass the local gravity test, and satisfy the stability conditions. To efficiently investigate cosmological tensions in $f(R)$ gravity, we consider the viable Hu-Sawicki $f(R)$ model (hereafter HS model) \cite{19} in this work and it is given by 
\begin{equation}
f(R)=-\frac{2\Lambda R^n}{R^n+\mu^{2n}}, \label{6}
\end{equation}  
where $\mu$ and $n$ are free parameters characterizing this model. By adopting $R\gg\mu^2$, the approximate $f(R)$ function shall be written as 
\begin{equation}
f(R)=-2\Lambda-\frac{f_{R0}}{n}\frac{R_0^{n+1}}{R^n}, \label{7}
\end{equation}
where $R_0$ is the present-day value of Ricci scalar and $f_{R0}=f_R(R_0)=-2\Lambda\mu^2/R_0^2$. For the purpose of constraining this model with data, one should first obtain the evolution of background and perturbation by inserting Eq.(\ref{7}) into Eqs.(\ref{3}-\ref{4}).

To the best of our knowledge, there are three main methods to confront $f(R)$ gravity with cosmological observations. The first is numerically solving the above equations in a direct way \cite{20,21,22,23,24,25,26,27,28,29}. The second is adopting an approximate framework to obtain the analytic solutions of the above equations and this method, to a large extent, can save computational cost \cite{30,31,32,33, a1}. The third one is studying the effects of viable $f(R)$ gravity on the large scale structure formation by using N-body and hydrodynamical simulations \cite{34}. Note that the last method always spend more computational cost and storage space than two previous ones.  

\begin{figure}[htbp] 
	\centering
	\includegraphics[scale=0.5]{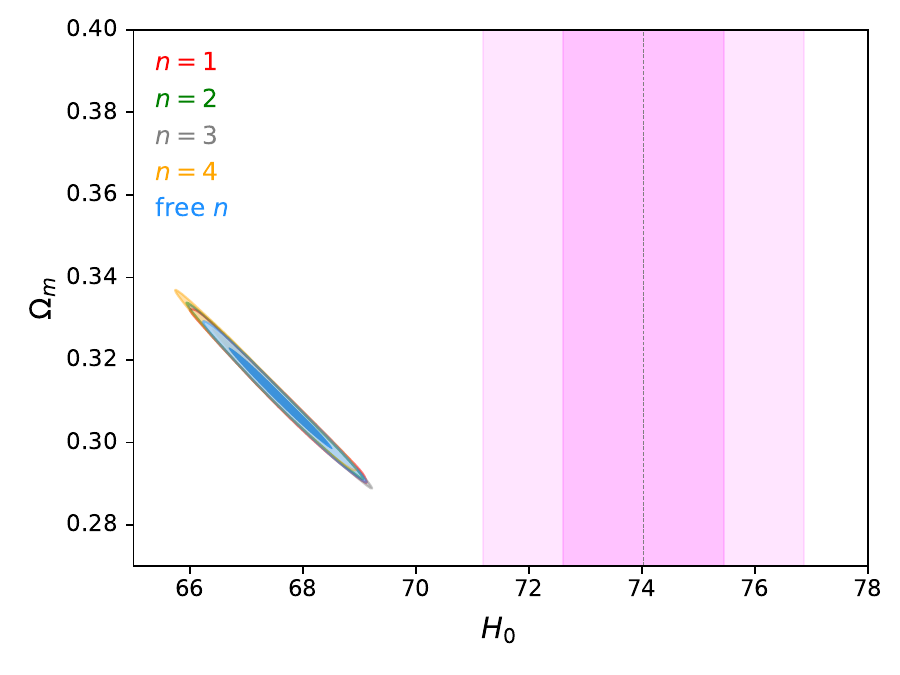}
	\hspace{10mm}
	\includegraphics[scale=0.5]{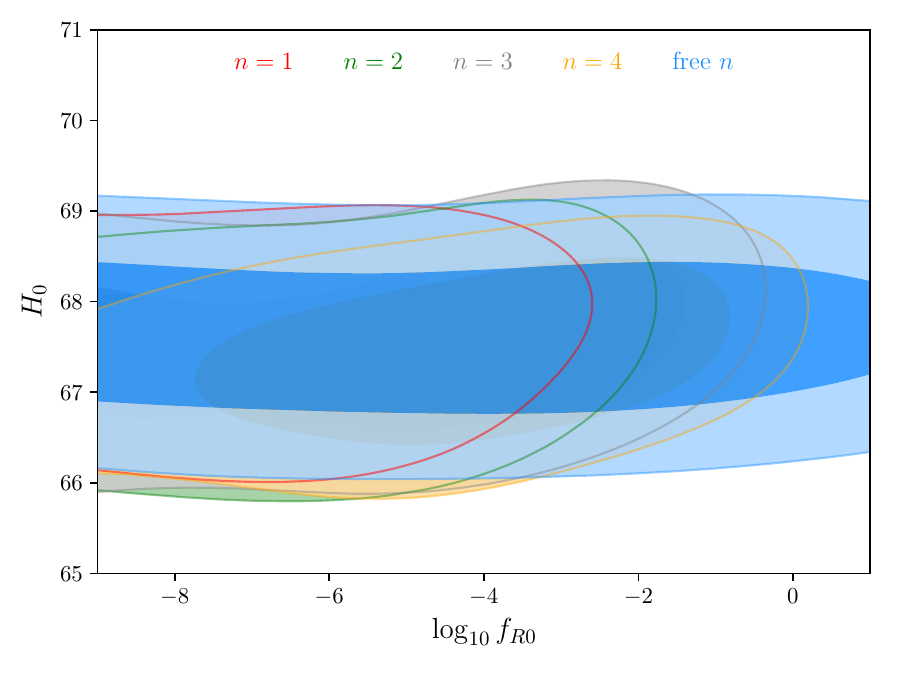}
	\caption{ The constrained 2-dimensional parameter spaces ($H_0$, $\Omega_{m0}$) and ($\log_{10}f_{R0}$, $H_0$) from the ``C'' dataset are shown for HS $f(R)$ models with $n=1$ (red), 2 (green), 3 (grey), 4 (orange) and free $n$ (blue), respectively. The grey dashed line and magenta bands denote $H_0=74.03\pm1.42$ km s$^{-1}$ Mpc$^{-1}$ measured by the HST \cite{14}.               }\label{f1}
\end{figure}

\begin{figure}[htbp] 
	\centering
	\includegraphics[scale=0.385]{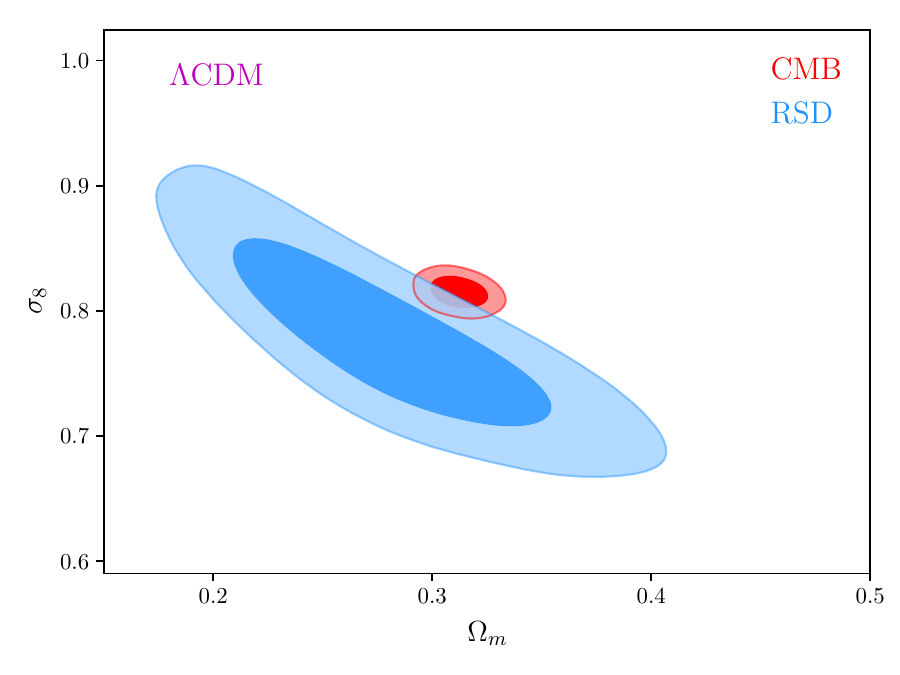}
	\includegraphics[scale=0.385]{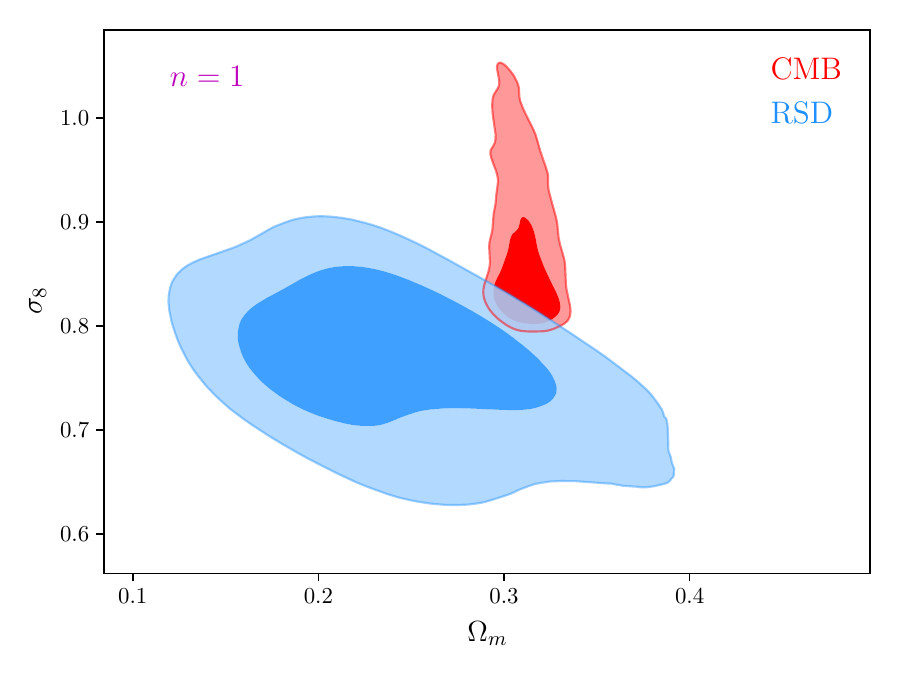}
	\includegraphics[scale=0.385]{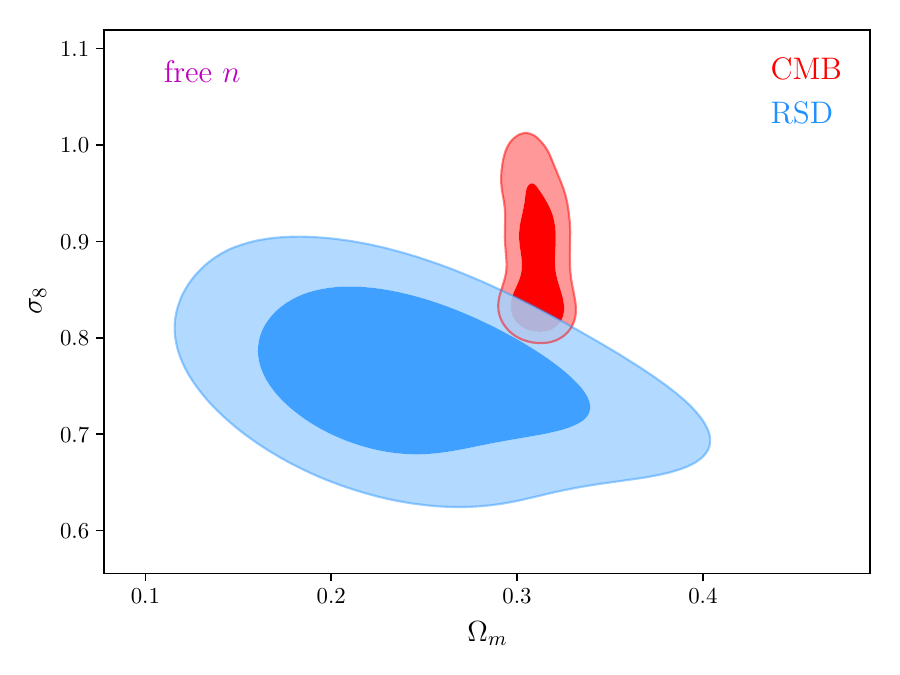}
	\caption{The constrained 2-dimensional parameter spaces ($\Omega_m$, $\sigma_8$) from the ``C'' (red) and ``R'' (blue) datasets are shown for $\Lambda$CDM, HS $f(R)$ models with $n=1$ and free $n$, respectively.  
	 }
	\label{f2}
\end{figure}

\begin{figure}[htbp] 
	\centering
	\includegraphics[scale=0.5]{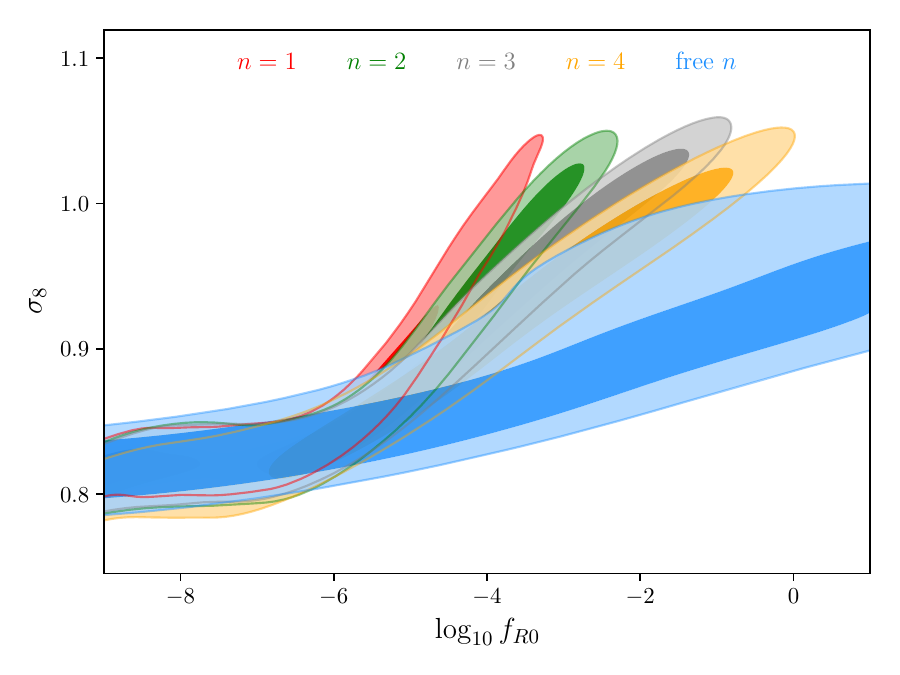}
	\caption{ The constrained 2-dimensional parameter spaces ($\log_{10}f_{R0}$, $\sigma_8$) from  the ``C'' dataset are shown for HS $f(R)$ models with $n=1$ (red), 2 (green), 3 (grey), 4 (orange) and free $n$ (blue), respectively.    }\label{f3}
\end{figure}

\begin{figure}[htbp] 
	\centering
	\includegraphics[scale=0.5]{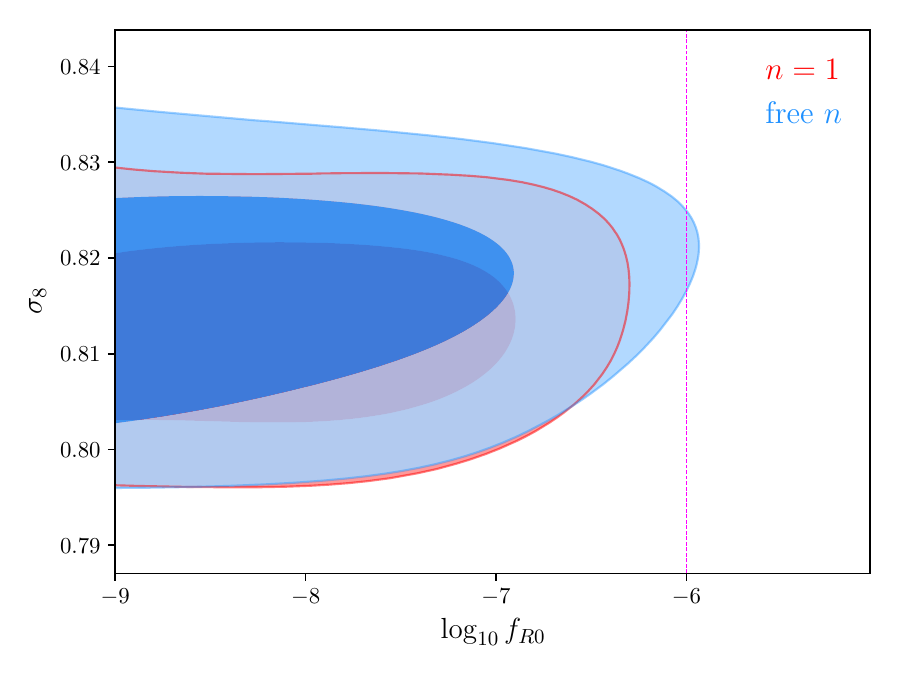}
	\caption{The constrained 2-dimensional parameter spaces ($\log_{10}f_{R0}$, $\sigma_8$) from  the data combination ``CBSHR'' are shown for HS $f(R)$ models with $n=1$ (red) and free $n$ (blue), respectively. The magenta dashed line denotes $\log_{10}f_{R0}=-6$. }\label{f4}
\end{figure}

\begin{figure}[htbp] 
	\centering
	\includegraphics[scale=0.5]{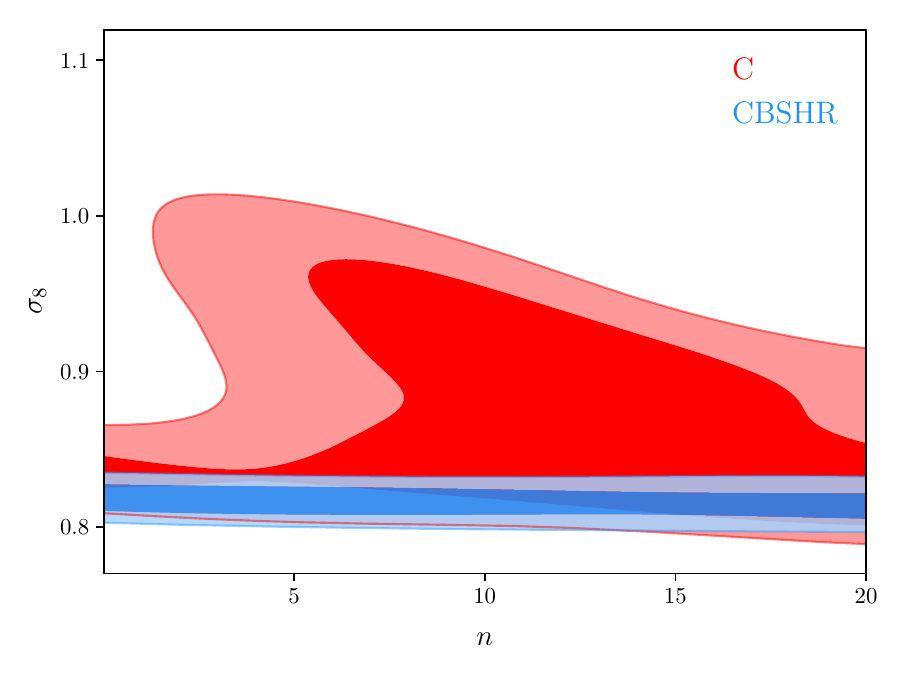}
	\caption{The constrained 2-dimensional parameter spaces ($n$, $\sigma_8$) are shown for HS $f(R)$ model with free $n$ by using the ``C'' (red) and ``CBSHR'' (blue) datasets, respectively. }\label{f5}
\end{figure}

\begin{figure}[htbp] 
	\centering
	\includegraphics[scale=0.465]{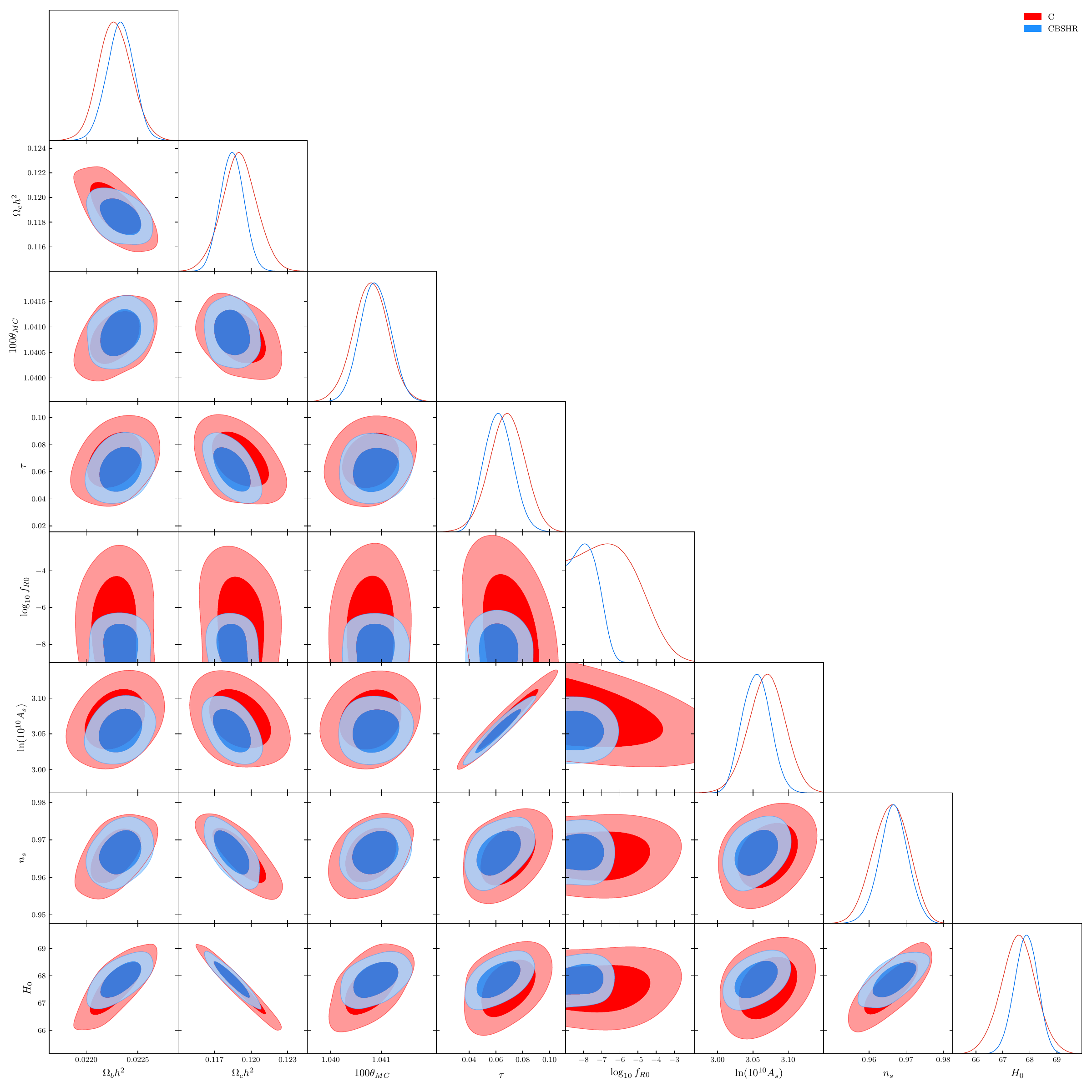}
	\caption{The marginalized constraints on the HS $f(R)$ model with $n=1$ are shown by using the  ``C'' (red) and ``CBSHR'' (blue) datasets, respectively.}\label{f6}
\end{figure}

\begin{figure}[htbp] 
	\centering
	\includegraphics[scale=0.41]{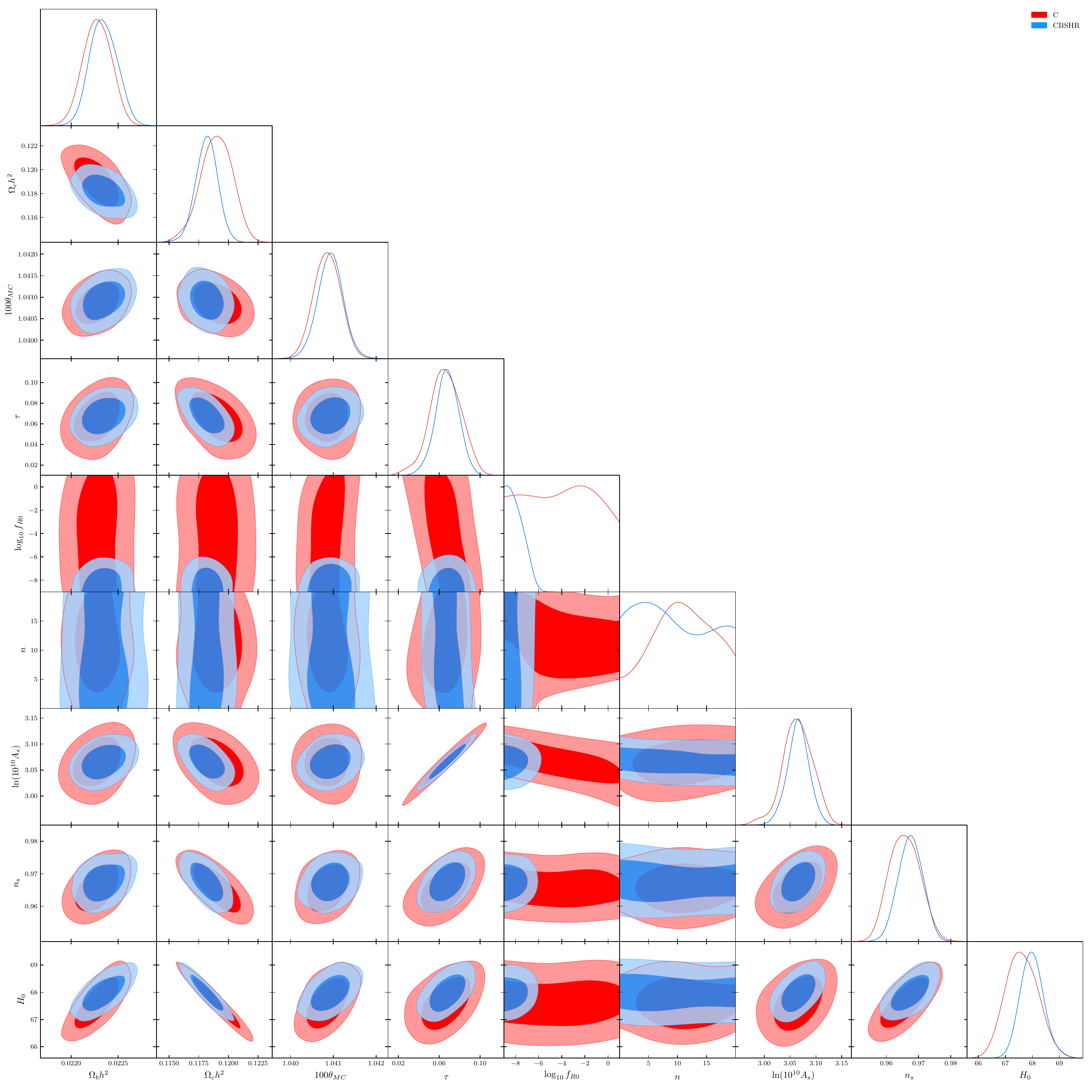}
	\caption{The marginalized constraints on the HS $f(R)$ model with free $n$ are shown by using the  ``C'' (red) and ``CBSHR'' (blue) datasets, respectively. }\label{f7}
\end{figure}

\section{Data and method}
Since our aim is to study whether HS $f(R)$ gravity can alleviate the $H_0$ and $\sigma_8$ tensions, first of all, we use the following two main datasets.

{\it CMB}: Although the mission of the Planck satellite is completed, its meaning for cosmology and astrophysics is extremely important. It has measured many aspects of formation and evolution of the universe such as matter components, topology and large scale structure effects. Here we  shall use updated Planck-2018 CMB temperature and polarization data including likelihoods of temperature at $30\leqslant \ell\leqslant 2500$ and the low-$\ell$ temperature and polarization likelihoods at $2\leqslant \ell\leqslant 29$, namely TTTEEE$+$lowE, and Planck-2018 CMB lensing data \cite{5}. We denote this dataset as ``C''.

{\it RSD}: To study the alleviation of $\sigma_8$ tension in $f(R)$ gravity, we adopt the redshift space distortions (RSD) as our reference probe which is sensitive to large scale structure formation. Specifically, we use the so-called ``Gold-2018'' growth-rate dataset \cite{35}. This dataset is denoted as ``R''.

Furthermore, to break the parameter degeneracy and give tight constraints on on free parameters of HS model, we also employ the following four probes.
  
{\it BAO}: By measuring the position of these oscillations in the matter power spectrum at different redshifts, the BAO, a standard cosmological ruler, can place constraints on the expansion history of the universe after decoupling and break the parameter degeneracy better. It is unaffected by errors in the nonlinear evolution of the matter density field and other systematic uncertainties. Specifically, we take the 6dFGS sample at the effective redshifts $z_{eff}=$ 0.106 \cite{36}, the SDSS-MGS one at $z_{eff}=$ 0.15 \cite{37} and the BOSS DR12 dataset at three effective redshifts $z_{eff}=$ 0.38, 0.51 and 0.61 \cite{38}. Specifically, to constrain the HS $f(R)$ gravity, we use the background quantity $D_A/r_d$ as a function of scale factor $a$ in the numerical analysis, where $D_A$ and $r_d$ are angular diameter distance and comoving BAO scale, respectively. To calculate the comoving sound horizon $r_d$, we use the fitting formula given by Ref.\cite{a2}.  This dataset is identified as ``B''. 

{\it SNe Ia}: SNe Ia, the so-called standard candle, is a powerful distance indicator to study the background evolution of the universe, particularly, the Hubble parameter and EoS of DE. In this analysis, we use the largest SNe Ia ``Pantheon'' sample today, which integrates the SNe Ia data from the Pan-STARRS1, SNLS, SDSS, low-z and HST surveys and encompasses 1048 spectroscopically confirmed points in the redshift range $z \in [0.01, 2.3]$ \cite{39}. In our numerical analysis, we use the full Pantheon sample and marginalize over the absolute magnitude parameter $M$. We refer to this dataset as ``S''.

{\it Cosmic Chronometers}: As a complementary probe to investigate the late-time evolution of the universe, we also include the cosmic chronometers in our numerical analysis. Specifically, we employ 30 chronometers to constrain the HS model \cite{40}. Hereafter we denote this dataset as ``H''.


It is worth noting that we take the first method (see Section II), namely numerically solving the background and perturbation equations, to implement constraints on HS $f(R)$ model. 
In order to obtain the posterior probability density distributions of model parameters, we incorporate the modified equations governing the evolution of background and perturbation of HS $f(R)$ model into the public online packages {\it CAMB} and {\it CosmoMC} \cite{41,42}. Specifically, we roughly calculate the Hubble expansion rate $H(a)$ and the linear growth factor $\delta(k,a)$ at each step of $a$, and use a interpolating scheme to obtain the solutions $H(a)$ and $\delta(k,a)$ with varying $a$.  As a consequence, we can numerically obtain the corresponding cosmological observables to be confronted with data. The latter package can be used for implementing a standard Bayesian analysis via the Markov Chain Monte Carlo (MCMC) method to infer the posterior probability density distributions of parameters. We use the Gelman-Rubin statistic $R-1=0.1$ as the convergence criterion of MCMC analysis. Meanwhile, to analyze the MCMC chains, we take the public package {\it GetDist} \cite{a3}. For HS $f(R)$ model, we choose the following prior ranges for different parameters:
$\Omega_bh^2 \in [0.005, 0.1]$, $\Omega_ch^2 \in [0.001, 0.99]$, $100\theta_{MC} \in [0.5, 10]$,  $\mathrm{ln}(10^{10}A_s) \in [2, 4]$, $n_s \in [0.8, 1.2]$ $\tau \in [0.01, 0.8]$, $\log_{10} f_{R,0} \in [-9, 1]$ and $n \in [0, 20]$.

To investigate comprehensively both $H_0$ and $\sigma_8$ tensions in HS model, we carry out the following numerical analysis. For $H_0$ tension, respectively, we constrain five models, i.e., $n=1, 2, 3, 4$ and free $n$ with the ``C'' dataset when keeping the typical parameter $\log_{10} f_{R,0}$ free. For $\sigma_8$ tension, we just present the constraining results of the representative case $n=1$ and the general one free $n$ from ``C'' and ``R'' datasets, respectively. We also display the comprehensive constraints on two models ($n=1$ and free $n$) by using the data combination ``CBSHR''. The corresponding $\chi^2$ expressions for all the datasets can be found in Ref.\cite{5}.

\begin{table*}[!t]
	\renewcommand\arraystretch{1.5}
	\caption{The marginalized constraints on the HS $f(R)$ models with $n=1$, 2, 3, 4 and free $n$ using the ``C'' dataset are shown, respectively.  For the typical parameter $\log_{10}f_{R0}$, we quote $2\sigma$ ($95\%$) uncertainties or bounds. The symbol ``$\diamondsuit$'' denotes the parameter that cannot be well constrained by observed data.}
	\begin{tabular} { l |c| c |c| c| c }
		\hline
		\hline
		
		Data               &   \multicolumn{5}{c}{C}                                        \\
		\hline
		Model              &  $n=1$      &$n=2$      &$ n=3  $     &$n=4 $    & free $n$    \\
		\hline
		{\boldmath$\Omega_b h^2   $} & $0.02228\pm 0.00016 $     & $0.02226\pm 0.00015        $    &$0.02228\pm 0.00016        $    & $0.02226\pm 0.00016        $  &$0.02228\pm 0.00015        $        \\
		
		{\boldmath$\Omega_c h^2   $} & $0.1190\pm 0.0014          $    & $0.1194\pm 0.0014          $   & $0.1190\pm 0.0014          $   &  $0.1195\pm 0.0015          $ & $0.1190\pm 0.0013          $                                                     \\
		
		{\boldmath$100\theta_{MC} $} & $1.04079\pm 0.00034        $   & $1.04079\pm 0.00031        $ & $1.04080^{+0.00032}_{-0.00029}$  & $1.04081\pm 0.00034        $ &   $1.04086\pm 0.00031        $                                      \\
		
		{\boldmath$\tau           $} & $0.068\pm 0.013            $     &$0.060^{+0.011}_{-0.015}   $  & $0.058\pm 0.016            $   & $0.056^{+0.019}_{-0.013}   $   & $0.066\pm 0.015            $                                \\
		
		{\boldmath${\rm{ln}}(10^{10} A_s)$} & $3.070\pm 0.024            $  & $3.055^{+0.021}_{-0.025}   $   &$3.048\pm 0.032            $   & $3.045^{+0.035}_{-0.026}   $ & $3.065^{+0.030}_{-0.027}   $                                       \\
		
		{\boldmath$n_s            $} & $0.9659\pm 0.0047          $    & $0.9648\pm 0.0045          $  & $0.9659\pm 0.0046          $  &$0.9645\pm 0.0050          $ & $0.9657\pm 0.0047          $                                       \\

		{\boldmath$\log_{10}  f_{R0}$}     & $ < -4.02$ (2$\sigma$)    & $ < -3.00$ (2$\sigma$)   & $ < -1.68$  (2$\sigma$)  & $-4.1^{+3.6}_{-4.3}$ (2$\sigma$) & $\diamondsuit$                                     \\

        {\boldmath$n$}  & ---    & ---   & ---    & --- & $\diamondsuit$                                                  \\
		
		\hline
		
		{\boldmath$H_0                       $ }& $67.58\pm 0.64             $ & $67.45\pm 0.63             $   & $67.59\pm 0.65             $    & $67.42\pm 0.67             $  & $67.61\pm 0.60             $                      \\
		
		{\boldmath$\Omega_m                  $ }& $0.3110\pm 0.0087          $  & $0.3128\pm 0.0087          $     & $0.3108\pm 0.0088          $   & $0.3134^{+0.0086}_{-0.010} $ & $0.3105\pm 0.0081          $                                                      \\
		
		{\boldmath$\sigma_8                  $ }& $0.859^{+0.041}_{-0.051} $  & $0.884^{+0.120}_{-0.081} $         & $0.908\pm 0.076            $ & $0.909\pm 0.068            $   &  $0.878^{+0.043}_{-0.065}   $                                                  \\
		\hline
		{\boldmath$\chi^2$    }                   & 12958.8                               &   12958.2                             &  12958.3                            &  12958.1                       &   12958.0 \\
		\hline
		\hline
	\end{tabular}
	\label{t1}
\end{table*}

\begin{table*}[!t]
	\renewcommand\arraystretch{1.5}
	\caption{The marginalized constraints on the HS $f(R)$ models with $n=1$ and free $n$  are shown by using the  ``R'' and ``CBSHR'' datasets, respectively. Similarly, for the typical parameter $\log_{10}f_{R0}$, we quote $2\sigma$ ($95\%$) uncertainties. The symbol ``$\diamondsuit$'' denotes the parameter that cannot be well constrained by observed data.         }
	\begin{tabular} { l |c| c |c| c }
		\hline
		\hline
		
		Data                 & \multicolumn{2}{c}{R}      &   \multicolumn{2}{|c}{CBSHR}                              \\
		\hline
		Model              &  $n=1$      & free $n$     & $n=1$  & free $n$  \\
		\hline
		{\boldmath$\Omega_b h^2   $} & ---     & ---                &$0.02233\pm 0.00013        $    & $0.02234\pm 0.00014        $      \\
		
		{\boldmath$\Omega_c h^2   $} & ---    & ---                 & $0.11844\pm 0.00095        $  & $0.11816\pm 0.00093        $                                      \\
		
		{\boldmath$100\theta_{MC} $} & ---   & ---                  & $1.04088\pm 0.00029        $  & $1.04093\pm 0.00030        $                                        \\
		
		{\boldmath$\tau           $} & ---     & ---                & $0.062\pm 0.010            $   & $0.068\pm 0.011            $                                   \\
		
		{\boldmath${\rm{ln}}(10^{10} A_s)$} & ---  & ---             & $3.055^{+0.019}_{-0.021}   $   & $3.066\pm 0.021            $                                        \\
		
		{\boldmath$n_s            $} & ---   & ---                   & $0.9666\pm 0.0038          $  & $0.9675\pm 0.0038          $                                        \\
		
		{\boldmath$\log_{10}  f_{R0}$}    & $< -0.773$ $(2\sigma)$    & $\diamondsuit$   & $< -6.75$ $(2\sigma)$    & $< -6.60$ $(2\sigma)$                                \\
		
		{\boldmath$n$}  & ---    & $\diamondsuit$  & ---   & $\diamondsuit$                                                                                                                          \\
		
		\hline
		
		{\boldmath$H_0                       $ }& $75\pm 10                  $ & $> 63.7$ $(2\sigma)$    & $67.86\pm 0.42             $    & $67.99^{+0.40}_{-0.45}     $                         \\
		
		{\boldmath$\Omega_m                  $ }& $0.243^{+0.044}_{-0.060}   $ & $0.245^{+0.045}_{-0.063}   $    & $0.3071\pm 0.0057          $  & $0.3054\pm 0.0056          $                                                    \\
		
		{\boldmath$\sigma_8                  $ }& $0.769^{+0.056}_{-0.043}   $  & $0.761\pm 0.055            $      & $0.8128\pm 0.0073          $   &  $0.823^{+0.0100}_{-0.0089}  $                                                    \\
		
		\hline
		{\boldmath  $\chi^2$ }          &      13.9                 &  13.1               &   14059.026            & 14059.023                                                                                \\

		\hline
		\hline
	\end{tabular}
	\label{t2}
\end{table*}

\section{Numerical Results}
For the purpose of studying the alleviation of two important cosmological tensions in the framework of HS $f(R)$ models, our main numerical results are displayed in Fig.\ref{f1} and Fig.\ref{f2} and marginalized constraining results are presented in Tab.\ref{t1} and Tab.\ref{t2}. We find that the $H_0$ values (see Tab.\ref{f1}) derived from Planck-18 data in five HS models are now $4.14\sigma$, $4.23\sigma$, $4.12\sigma$, $4.21\sigma$ and $4.16\sigma$ lower than that directly measured by the HST, and that these new $H_0$ values relieve hardly the existing $4.39\sigma$ tension under the assumption of $\Lambda$CDM. This implies that HS gravity behaves very similar to $\Lambda$CDM at the background level. To some extent, one can predict its $H_0$ behavior via Eq.(\ref{7}). In Fig.\ref{f1}, we have exhibited the constrained 2-dimensional parameter spaces ($\Omega_m$, $\sigma_8$) for five HS models, it is easy to see the large $H_0$ gap between CMB and HST observations. Only using the CMB data, we conclude that $H_0$ is insensitive to typical model parameter $\log_{10}f_{R0}$ in all five models (see the right panel of Fig.\ref{f1}). To investigate the $\sigma_8$ tension, in Fig.\ref{f2}, first of all, we display the constrained $\Omega_m$-$\sigma_8$ plane for $\Lambda$CDM as a reference. Then, we present constrained $\Omega_m$-$\sigma_8$ planes for the commonly used HS model with $n=1$ and for the complete HS model with free $n$, respectively. We find that the relatively small $\sigma_8$ discrepancy in both considered HS scenarios with a little larger parameter spaces ($\Omega_m$, $\sigma_8$) cannot be resolved and is still over $1\sigma$ level. This implies that the HS $f(R)$ gravity cannot reduce current $H_0$ and $\sigma_8$ tensions, which is the key result of this work.       

It is also interesting to study the parameter degeneracy between $\log_{10}f_{R0}$ and $\sigma_8$. When only using CMB data, for five HS models, one may find that $\log_{10}f_{R0}$ is positively correlated with $\sigma_8$, which indicates that stronger deviations in HS f(R) gravity from GR lead to larger effects of matter clustering (see Fig.\ref{f3}). However, when using the combined dataset CBSHR, this positive correlation disappears and $\sigma_8$ seems to be insensitive to $\log_{10}f_{R0}$ (see Fig.\ref{f4}). Meanwhile, we are also of interests to study the degeneracy between the additional parameter $n$ and $\sigma_8$, and find that the amplitude of matter clustering is insensitive to $n$ regardless of the use of C or CBSHR datasets (see Fig.\ref{f5}). Furthermore, to study the degeneracies between parameters better, we exhibit the marginalized constraints on HS $f(R)$ models with $n=1$ and free $n$ in Fig.\ref{f6} and Fig.\ref{f7}, and obtain the following conclusions: (i) to a large extent, the parameter spaces are compressed when combining C with BSHR datasets; (ii) in all cases, two typical parameters $\log_{10}f_{R0}$ and $n$ are insensitive to other cosmological parameters, which is clarified for the first time in the literature.  

We also find that when using only CMB data, the case of free $n$ has the smallest $\chi^2=12958.0$ but close to other ones, when using only RSD data, the case of free has a relatively better fitting than HS model with $n=1$, and that when using the combined datasets CBSHR, these two cases present almost same $\chi^2$ value. Therefore, we can not easily distinguish these HS $f(R)$ variants from currently statistical analysis.     

Moreover, in Tab.\ref{t1}, we can find that the best constraint $\log_{10}f_{R0} < -4.02$ at the $2\sigma$ confidence level originates  from the case of $n=1$ by only using CMB data, while two typical parameters $\log_{10}f_{R0}$ and $n$ in the free $n$ case cannot be well constrained (see also Fig.\ref{f6} and Fig.\ref{f7}). Subsequently, in Tab.\ref{t2}, we find that,  when using RSD data alone, constraints on typical parameters of HS models are poor and smaller $\sigma_8$ values are obtained, which indicates that this RSD dataset gives a smaller effect of matter clustering at late times than the CMB observation. Interestingly, although the mean value of the constraint $H_0=75\pm10$ km s$^{-1}$ Mpc$^{-1}$ from RSD data is consistent with the HST result, it has a much larger uncertainty. Finally,  at the $2\sigma$ confidence level, we give our best constraint on the typical parameter $\log_{10}f_{R0}<-6.75$ in the case of $n=1$, while $\log_{10}f_{R0}<-6.60$ in the free $n$ case. It is worth noting that we still cannot provide good constraint on $n$ even using the joint dataset CBSHR. 

It is noteworthy that there are two interesting and tight constraints from large scale structure observations. 
In Ref.\cite{21}, the authors uses the galaxy clustering ratio, a sensitive probe of the nature of gravity in the cosmological regime, gives ${f_{R0}}< 4.6\times 10^{-5}$ at the $2\sigma$ level. Recently, in Ref.\cite{29}, the authors place constraints on chameleon-$f(R)$ gravity from galaxy rotation curves and find that $f(R)$ models within the range $-7.5<\log_{10}f_{R0}<-6.5$ seem to be favored with respect to $\Lambda$CDM. Interestingly, our best constraint just lies in this range and this may give a clue of the correct living range for the HS $f(R)$ gravity.

\section{Discussions and conclusions}  
Recently, the $H_0$ and $\sigma_8$ tensions under the standard cosmological paradigm have re-activated a wide variety of alternative cosmological models. However, all the time, there is a lack of direct tests of $f(R)$ gravity in resolving both tensions. To address this urgent issue, we confront the popular HS $f(R)$ gravity with current observations. By testing five specific HS $f(R)$ models with observational datasets, we obtain two main conclusions: (i) HS $f(R)$ gravity cannot resolve both $H_0$ and $\sigma_8$ tensions; (i) the typical parameters $\log_{10}f_{R0}$ and $n$ are insensitive to other cosmological parameters. Meanwhile, in the HS $f(R)$ model with $n=1$, we give our best constraint $\log_{10}f_{R0}<-6.75$ at the $2\sigma$ confidence level. 

It is noteworthy that a coupling between matter and geometry in the framework of $f(R)$ gravity may help resolve these tensions, and that other $f(R)$ gravity models may relieve both discrepancies much better than the considered HS $f(R)$ one. We expect that future high-precision CMB and SNe Ia observations and independent probes such as gravitational waves could help reduce or even solve these intractable cosmological tensions.  

\section{Acknowledgements}
Deng Wang warmly thank the anonymous referee for helpful comments on this manuscript, David Mota for useful discussions on modified gravity and Yuan Sun for useful communications on gravitational theories.  This work is supported by the Ministry of Science and Technology of China under Grant No.2017YFB0203300, National Nature Science Foundation of China under Grants No.11988101 and No.11851301.

\end{document}